\documentclass[aps,prd,showpacs,nofootinbib,floats,floatfix,preprintnumbers,groupedaddress,twocolumn]{revtex4}

\usepackage{bm}
\usepackage{latexsym}
\usepackage{dcolumn}
\usepackage{amsmath,amsfonts,amssymb}
\usepackage{graphicx,epsfig}
\usepackage{amsmath}
\usepackage{fancyhdr}
\usepackage{hyperref}
\usepackage{graphicx,epstopdf}
\hypersetup{
	colorlinks   = true, 
	urlcolor     = blue, 
	linkcolor    = blue, 
	citecolor   = red 
}

\begin{document}
\title{Thermogeometric description of the van der Waals like phase transition in AdS black holes}
\author{Krishnakanta Bhattacharya\footnote {\color{blue} krishnakanta@iitg.ernet.in}}
\author{Bibhas Ranjan Majhi\footnote {\color{blue} bibhas.majhi@iitg.ernet.in}}

\affiliation{Department of Physics, Indian Institute of Technology Guwahati, Guwahati 781039, Assam, India}

\date{\today}

\begin{abstract}
It is well known that interpreting the cosmological constant as the pressure, the AdS black holes behave as the van der Waals thermodynamic system. In this case, like a phase transition from vapor to liquid in a usual van der Waals system, black holes also changes phases about a critical point in the $P$-$V$ picture, where $P$ is the pressure and $V$ is the thermodynamic volume. Here, we give a geometrical description of this phase transition. Defining the relevant Legendre invariant thermogeometrics corresponding to the two criticality conditions, which determine the critical values of respective thermodynamical entities, we show that the critical point refers to the divergence of the Ricci scalars calculated from these metrics. The similar descriptions are also provided for the other two pictures of the van der Waals like phase transition: one is $T$-$S$ and the other one is $Y$-$X$ where $T$, $S$, $X$ and $Y$ are temperature, entropy, generalized force and generalized displacement; i.e. potential corresponding to external charge, respectively. The whole discussion is very general as no specific black hole metric is being used.  
\end{abstract}


\maketitle

\section{Introduction}
The pioneering works of Bakenstein \cite{Bekenstein:1973ur} and Hawking \cite{Hawking:1974sw, Bardeen:1973gs} first exhibited the thermodynamic property of the black holes which was the stepping stone of today's well-established conviction that the black hole horizons have thermodynamic structure in any gravitational theory. Later various properties of the black holes were studied in the thermodynamical way. To be on a per with the conventional thermodynamics, the phase transition in the black hole system is being explored since long ago, commenced by Davis \cite{Davies:1989ey} and later by Hawking and Page \cite{Hawking:1982dh} in AdS background. The results were extended further for the charged black holes with the analogy with the van der Waals fluid system \cite{Chamblin:1999tk, Chamblin:1999hg}. However, these works were unable to formulate the proper definition of pressure and the volume to investigate the P-V criticality of a black hole precisely. It must be mentioned that one can find the identification of the inverse of the horizon temperature and the radius of the black hole horizon as the pressure and the volume respectively in these earlier works. 

The anomaly was solved later only when the cosmological constant was endowed with the role of the pressure with a modified version of the first law \cite{Kastor:2009wy, Dolan:2010ha, Dolan:2011xt, Dolan:2011jm, Dolan:2012jh} where the black hole mass is identified as the enthalpy on the extended phase space. Recently, the introduction of the pressure term has been extended for the cases with quintessence, non-linear electromagnetism etc. \cite{ Azreg-Ainou:2014lua, Azreg-Ainou:2014twa}. While studying the phase transition of the black holes two approaches caught the attention of the physicists: one approach deals with the divergence of the heat capacity and the inverse of the isothermal compressibility \cite{Banerjee:2011cz, Banerjee:2012zm, Majhi:2012fz, Lala:2012jp, Ma:2014tka, Azreg-Ainou:2014gja, Liu:2013koa} and the other one discusses the phase transition of black holes in the AdS background with the cosmological constant being treated as the pressure. In the latter description the whole black hole system has been mapped to the van der Waals fluid system successfully in many avenues  with the fact that the $P$-$V$ diagram of the black holes at constant temperature and charges in the extended phase space looks exactly similar to the $P$-$V$ diagram of the isotherms of the van der Waals gas  system in the conventional thermodynamics. Using the criticality conditions, the critical exponents of the black hole phase transition are found to be those of the usual van der waals system \cite{Kubiznak:2012wp} (See also \cite{Kubiznak:2016qmn} for a recent review and list of all relevant references on this topic){\footnote{Very recently, $P$-$V$ criticallity has also been observed from the holographic concept \cite{Momeni:2016qfv}.}}. Recently, these two types of phase transition have been described in a general framework by one of the authors \cite{Mandal:2016anc, Majhi:2016txt} which shows that the values of the critical exponents are universal, not restricted to particular black hole. 

The phase transition, characterized by the divergence of the heat capacity, can be described in terms of geometry. This geometrical interpretation of the thermodynamic properties of the black holes was first introduced by Weinhold \cite{WEINHOLD} who defined a metric to describe the thermal property in a geometrical way. Later Ruppeiner \cite{RUPP, Ruppeiner:1995zz} developed another metric which is conformally equivalent to the Weinhold's metric where the inverse of the temperature is the conformal factor. The underlying idea of these works were to interpret the thermal interaction in terms of the curvature of the geometry. But, these two earlier metric formalism are not consistent to each other. Recently, Quevedo et al. has claimed that those inconsistency appears due to the fact that those metrics are not Legendre invariant and they have shown how to formulate the thermogeometrical metric in a Legendre-invariant way \cite{Quevedo:2006xk, Quevedo:2007mj, Quevedo:2008xn, Quevedo:2008ry, Alvarez:2008wa, Quevedo:2011np, Quevedo:2017tgz}. This have been further extended for the charged phantom AdS black holes \cite{Quevedo:2016cge, Quevedo:2016swn}. Therefore, from the analysis of the criticality in terms of the heat capacity, one can define a thermogeometrical metric in a Legendre invariant way such that the scalar curvature (Ricci scalar) diverges where the phase transition occurs, thereby interpreting the phase transition point as one with the diverging scalar curvature of the thermogeometrical metric. This method has been generalized more convincingly in a recent work by one of the authors \cite{Banerjee:2016nse} where no explicit black hole metric has been used. Here, it has been argued that the divergence of the Ricci scalar for the relevant Legendre invariant thermogeometrical metric can be the signature of the true phase transition.

Our aim of this paper is to give the geometric interpretation of the $P$-$V$ criticality in a general way; i.e. in a metric independent way. The idea is similar to the geometrical interpretation of phase transition defined in terms of the inverse of the heat capacity. As we know in $P-V$ picture, the pressure is expressed as a function of volume ($V$), temperature ($T$) and other external charges, denoted here by $Y_i$. Then for constant $T$ and $Y_i$, the critical point is defined by the two simultaneous conditions $(\partial P/\partial V)_{T,Y_i}=0$ and $(\partial^2 P/\partial V^2)_{T,Y_i}=0$: one corresponds to the vanishing of the tangent of curves in the $P-V$ diagram and the other one refers to the point of inflection. Use of these two leads to the critical values of the respective thermodynamical entities. Our precise goal is to describe these two conditions in terms of the thermogeometry in a covariant way. Since, the conditions are completely independent here we need two thermogeometrical metrics to describe them. We shall construct the explicit expressions of the thermogeometrics and show that for each condition the corresponding Ricci scalar diverges at the critical point. Here, the metrics will be constructed in a Lagendre invariant way borrowing the original idea presented in \cite{Quevedo:2006xk, Alvarez:2008wa}. 

It should be mentioned that besides $P-V$ criticality there are also $T-S$ \cite{Shen:2005nu, Zhang:2015ova, Zeng:2015tfj, Zeng:2015wtt, Spallucci:2013osa, Mo:2016apo, Kuang:2016caz} and $Y-X$ criticalities \cite{Ma:2016aat} for the AdS black holes where $Y$ refers to the charge corresponding to the potential $X$. It has been observed that the AdS black holes can behave like van der waals system when one looks at the thermodynamics either in $T-S$ picture or in $Y-X$ picture. In these cases the usual first law of thermodynamics is useful; i.e. the cosmological constant remains a true constant and we do not have to go in the extended phase space formalism. In this paper we also describe these cases in terms of thermogeometry. In this regard, it should be pointed out that there are some sporadic attempts \cite{Shen:2005nu,Zhang:2015ova,Mo:2016apo, Hendi:2015eca, Hendi:2015rja} in the $T-S$ picture as well as in the $Y-X$ picture \cite{Shen:2005nu}; but none of them are complete in the sense that both the conditions were not being dealt with and were confined to a particular black hole metric. Here we shall give a complete description of all the existing pictures without invoking any spacetime. Therefore our analysis is much more robust and very general.

   We organize our paper as follows. In the following section we shall describe the P-V criticality, constructing a pair of thermodynamic metrics in a Legendre invariant way and then showing that the two mentioned condition of the critical point will imply the divergence of the Ricci-scalars of the two thermodynamic metrics. On the following two sections, it will be examined whether one can give the geometric interpretation in the $T-S$ and $Y-X$ criticalities in the same way or not. It will be shown that one can really give geometric description in the same way for these two criticalities as well. Thereafter, the conclusion of our analysis will be given. To make our paper self-sufficient and, since, there is a the lack of complete derivation, we shall give the proof of the first law in the appendix for the AdS black holes with varying cosmological constant using the Wald's formalism (i.e. Lagrangian prescription). 
   
{\it Notations:} $P$, $V$, $T$ and $S$ refer to the pressure, thermodynamic volume, temperature and entropy, respectively. We denote $X_i$ as the potential corresponding to the charge $Y_i$. For example $X_i$ can be electromagnetic potential, angular velocity, etc whereas $Y_i$ includes electric charge, angular momentum, etc. 
\section{$P$-$V$ criticality in thermogeometric picture}
In the usual thermodynamics, the critical point is the extreme point of the vaporization curve (in the $P$-$T$ pahse diagram of the van der Waals gas). The gas above the critical temperature cannot be liquefied and, therefore, only one state exists which is the gaseous state. Mathematically, the critical point of the van der Waals gas system represents a particular point in the $P$-$V$ diagram of the critical isotherm where the tangent vanishes as well as the maxima and the minima of the $P$ vs $V$ curve meet to form a inflection point. Therefore, at the critical point the two independent conditions coincide; i.e. $P_V=0=P_{VV}$.

The same discussions can be applied for the black holes as well because it has been found that the $P$-$V$ diagram of the AdS black holes at constant temperature and charges looks identical to that for van der Waals gas system when one interprets the cosmological constant as pressure ($P$) (See \cite{Kubiznak:2012wp} and for the review see \cite{Kubiznak:2016qmn}).  Consequently, the critical conditions are obtained by setting $(\partial P/\partial V)_{T, Y_i}=0=(\partial^2 P/\partial V^2)_{T, Y_i}$. In this section, we want to obtain the geometrical interpretation of those conditions in terms of the thermogeometrical metric. We have already mentioned that the critical point refers to the singular behavior of the scalar curvature of the thermogeometrical metric. Here, our aim is to investigate such possibility for the $P$-$V$ criticality of the AdS black holes. Let us now find out the relevant thermogeometrical metric in a Legendre invariant way. The approach here will be adopted similar to \cite{Quevedo:2006xk, Alvarez:2008wa}.

\subsection{The first condition: $(\frac{\partial P}{\partial V})_{T,Y_i}=0$} \label{SEC1ST}
For the first condition, we found out the proper thermodynamic quantity to start with is the Helmholtz free energy $F$ which is the function of $V$, $T$ and $Y_i$ as shown later. The underlying motivation of choosing the Helmholtz free energy as the appropriate quantity will be understood from the main analysis. Since in this case the black hole mass $M$ is identified as the enthalpy  \cite{Kastor:2009wy}, the internal energy is given by $E=M-PV$ and the free energy is defined as 
\begin{equation}
F=E-TS=M-PV-TS~.
\label{F}
\end{equation}
The first law of the black holes under the AdS background with the role of the cosmological constant as the pressure  is given as 
\begin{align}
dM=TdS+VdP+\sum_{i}X_idY_i~. \label{1ST}
\end{align}
In the maximum cases, this has been obtained by explicit use of a black hole metric. Also there is a Hamiltonian approach \cite{Kastor:2009wy, Baldiotti:2017ywq} for general cases. In the appendix \ref{APPEN1}, we have given the Lagrangian analysis to derive the first law of an AdS black hole in a metric independent way. Although, we found an earlier work in this regards \cite{Couch:2016exn}, we discuss it in a more rigorous way and to do this we adopt the Wald's formalism \cite{Wald:1993nt} in absence of any hair. Inclusion of the existing hairs in the black hole is very standard which leads to the expression \eqref{1ST}. Using this and (\ref{F}) one gets
\begin{align}
dF=-PdV-SdT+\sum_iX_idY_i~,
\label{HEL1}
\end{align}
i.e. $F$ is function of $V$, $T$ and $Y_i$. Therefore, the conjugate quantities corresponding to these variables are
\begin{eqnarray}
&&P=-\Big(\frac{\partial F}{\partial V}\Big)_{T,Y_i}\equiv-F_V; \,\,\
S=-\Big(\frac{\partial F}{\partial T}\Big)_{V,Y_i}\equiv-F_T;
\nonumber
\\
&&X_i=\Big(\frac{\partial F}{\partial Y_i}\Big)_{T,V}\equiv F_{Y_i}~.
\label{PSX}
\end{eqnarray}
Note, in this case our thermodynamic potential is the function of both the extensive variables ($V$, $Y_i$) and the intensive variable (T). We define a thermodynamic phase space $\mathcal{T}$ with the coordinates $Z^A=\{F, \tilde{X}^a, \tilde{P}^a\}$ where $\tilde{X}^a=\{V, T, Y_i\}$ are the thermodynamic variables and $\tilde{P}^a=\{F_V=-P, F_T=-S, F_{Y_i}=X_i \}$ are the the conjugate variables. Note, the conjugate variables are the functions of the thermodynamic variables $\tilde{X}^a$. The approach which we shall follow here is developed by Hermann \cite{Hermann} and Mrugala \cite{Mrugala1, Mrugala2}, and it was followed later by Quevedo \cite{Quevedo:2006xk, Quevedo:2007mj, Quevedo:2008xn, Quevedo:2008ry, Alvarez:2008wa, Quevedo:2011np, Quevedo:2017tgz} extensively. We identify the fundamental one form on the phase space $\mathcal{T}$ in $F$ representation as 
\begin{align}
\theta_F=dF-\sum_{ab}\delta_{ab}\tilde{P}^ad\tilde{X}^b
\nonumber
\\
=dF+PdV+SdT-\sum_{i}X_idY_i~, \label{THF}
\end{align}
with $\delta_{ab}$=diag(1,...,1). If one defines a thermogeometrical metric $G_1^{(PV)}(Z^A)$ on $\mathcal{T}$, then the set ($\mathcal{T}$, $\theta_F$, $G_1^{(PV)}$) would define a contact Riemannian manifold when it satisfies $\theta_F\wedge(d\theta_F)^3\neq0$, where the operation $\wedge$ means the exterior product and $(d\theta_F)^3=d\theta_F\wedge d\theta_F\wedge d\theta_F$.
 Let there be a subspace $\mathcal{E}$, the coordinates of which are $\tilde{X}^a$. We assume there exists a smooth mapping $\varphi_F:\mathcal{E}\rightarrow\mathcal{T}$. The subspace $\mathcal{E}$ is called the space of equilibrium states if $\varphi^{*}_F(\theta_F)=0$. It means each thermodynamic system in equilibrium possess its own space $\mathcal{E}$. The fundamental Gibbs 1-form of \eqref{THF}, when projected on $\mathcal{E}$ with the pullback of $\varphi$, generates the first law of thermodynamics and the conditions for thermodynamic equilibrium. Note, while single charge is present in the theory, $\mathcal{T}$ and $\mathcal{E}$ is seven- and three-dimensional respectively. But, when multiple charges are present, the dimension of $\mathcal{T}$ and $\mathcal{E}$ depends on the number of charges.
 
To give the geometrical interpretation one has to define a metric properly. The recent trend of developing a thermogeometrical metric is to define the metric in a Legendre invariant way for obtaining physically consistent result. In our case, the Legendre transformations are defined as
\begin{eqnarray}
&&F_{old}=F_{new}-\delta_{ab}\tilde{X}^a_{new}\tilde{P}^b_{new};
\nonumber
\\
&&\tilde{X}^a_{old}=-\tilde{P}^a_{new}; \,\,\,\
\tilde{P}^a_{old}=\tilde{X}^a_{new}~.
\label{LT}
\end{eqnarray}

Now, one can choose a Legendre invariant metric in many ways. One possible form can be:
\begin{align}
G_1^{(PV)}=\Big(dF-\sum_{ab}\delta_{ab}\tilde{P}^ad\tilde{X}^b\Big)^2
\nonumber
\\
+\lambda\Big(\sum_{ab}\xi_{ab}\tilde{P}^a\tilde{X}^b\Big)\Big(\sum_{cd}\eta_{cd}d\tilde{P}^cd\tilde{X}^d\Big)~, \label{GENERAL}
\end{align}
where $\eta_{cd}=$diag(-1,...,1), $\lambda$ is arbitrary Legendre invariant function of $\tilde{X}^a$ and $\xi_{ab}$ is arbitrary constant diagonal function. Here, for the shake of simplicity, we take $\lambda=1$, and $\xi_{ab}$=diag(1,...,1). Thus, the simplified form can be written as
\begin{align}
G_1^{(PV)}=\theta_F^2+(-PV-ST+\sum_iX_iY_i)(dPdV
\nonumber
\\
-dSdT+ \sum_idX_idY_i)~. 
\label{new1}
\end{align}
 Note, while forming the metric we multiply the variables with their conjugate ones to get the dimensionally consistent result. One can check the above metric is Legendre invariant from the definition of the Legendre transformation \eqref{LT}.
The metric when induced on $\mathcal{E}$ by means of $\mathcal{G}_1^{(PV)}=\varphi_F^{*}(G_1^{(PV)})$, it yields
 \begin{align}
\mathcal{G}_1^{(PV)}=(-PV-ST+\sum_iX_iY_i)[-F_{VV}dV^2+F_{TT}dT^2
 \nonumber
 \\
 +\sum_{i,j}F_{Y_iY_j}dY_idY_j+2\sum_iF_{TY_i}dTdY_i]~,
 \label{new2}
 \end{align}
where in the above we use the fact that the conjugate variables are the functions of the thermodynamic variables $\tilde{X}^a$. Now, one can verify that the exact nature of divergence of the Ricci scalar on $F_{VV}$ turns out to be
\begin{equation}
R_1^{(PV)}|_{max. diver.} \sim \mathcal{O}(\frac{1}{F_{VV}^2})~.
\label{R}
\end{equation}
This can be observed from the explicit expression of the Ricci-scalar for a single charged metric. Although, for the presence of the multiple charges, the exact form of it can not be given but, the nature is exactly the same like the simple one. For the presence of the single charge, the expression of the scalar curvature is explicitly provided in the Appendix \ref{APPEN2} (see \eqref{RM}). 
The metric coefficients can be identified as 
\begin{eqnarray}
&&f(V,T,Y)=F_{VV}(VF_{V}+TF_T+YF_Y)~; 
\nonumber
\\
&&g(V,T,Y)=F_{TT}(VF_{V}+TF_T+YF_Y)~;
\nonumber
\\
&&A(V,T,Y)=F_{YY}(VF_{V}+TF_T+YF_Y)~; 
\nonumber
\\
&&h(V,T,Y)=F_{TY}(VF_{V}+TF_T+YF_Y)~. \label{COEFF}
\end{eqnarray}
Now the expression (\ref{RM}) tells that the Ricci scalar diverges as $\mathcal{O}(1/f^2)$.  As $f$ is given by the first equation of \eqref{COEFF}, one obtains \eqref{R}.
Since $P$ is given by the first relation of (\ref{PSX}), we have $\partial P/\partial V = -F_{VV}$. Therefore, one of the two conditions of determining the critical point of the phase transition leads to the vanishing of $F_{VV}$ and, consequently, the calculated Ricci-scalar diverges. Later, we shall give an explicit example to make more elaborative comments on these discussions. Therefore, we conclude that the vanishing of the tangent condition is equivalent to the divergence of the Ricci scalar of the thermogeometry in $F$ picture and this is an invariant statement as the quantity is a scalar.  In the below we shall concentrate on the other condition. 

Before going to the next analysis, let us mention that the thermodynamic metrics for the later purposes will be constructed in a similar manner as has been done in this section. We shall follow the definition of the metric mentioned in \eqref{GENERAL} with proper identification of the variables $\tilde{X}^a$, corresponding conjugate quantities $\tilde{P}^a$ and proper thermodynamic quantity (instead of $F$) along with the constraint \eqref{LT} to construct it in a Legendre invariant way. 

\subsection{The second condition: $(\frac{\partial^2 P}{\partial V^2})_{T,Y_i}=0$} \label{SEC2ND}
To get the geometrical interpretation of the second condition (point of inflection) let us take pressure as a function of the volume, the temperature and the charges, i.e., $P=P(V,T,Y_i)$.  It implies
\begin{align}
dP=\Big(\frac{\partial P}{\partial V}\Big)_{T,Y_i}dV+\Big(\frac{\partial P}{\partial T}\Big)_{V,Y_i}dT+\sum_i\Big(\frac{\partial P}{\partial Y_i}\Big)_{V,T}dY_i
\nonumber
\\
=P_VdV+P_TdT+\sum_iP_{Y_i}dY_i~. 
\label{DP}
\end{align}
Here also we define the thermodynamic phase space $\mathcal{T}$ with the coordinates $Z^A=\{P, \tilde{X}^a, \tilde{P}^a\}$ where $\tilde{X}^a=\{V, T, Y_i\}$ are the variables and $\tilde{P}^a=\{P_V, P_T, P_{Y_i}\}$ are the corresponding conjugate quantities of those variables.
As done earlier, we again choose a metric properly in the $\theta_P$ invariant form with $\theta_P=dP-P_VdV-P_TdT-\sum_iP_{Y_i}dY_i$. Again we consider the subspace $\mathcal{E}$ having the coordinates $\tilde{X}^a$. Also we assume a smooth mapping $\varphi_P:\mathcal{E}\rightarrow\mathcal{T}$ and the subspace $\mathcal{E}$ is called the space of equilibrium states if $\varphi^{*}_P(\theta_P)=0$.  Let us take the same definition of \eqref{GENERAL} to form the metric with $F$ being replaced by $P$. In this case which yields:
\begin{align}
G_2^{(PV)}=\theta_P^2+(VP_V+TP_T+\sum_iY_iP_{Y_i})(-dVdP_V
\nonumber
\\
+dTdP_T+\sum_jdY_jdP_{Y_j}). \label{THP}
\end{align}
The metric is Legendre invariant according to the definition \eqref{LT} with $F$ being replaced by $P$. For $\mathcal{G}_2^{(PV)}=\varphi_P^{*}(G_2^{(PV)})$ one obtains
\begin{align}
\mathcal{G}_2^{(PV)}
=(VP_V+TP_T+\sum_iY_iP_{Y_i})(-P_{VV}dV^2
\nonumber
\\
+P_{TT}dT^2+\sum_{j,k}P_{Y_jY_k}dY_jdY_k+2\sum_lP_{TY_l}dTdY_l)~.
\label{G2}
\end{align}
Again, to obtain the above equation, one has to use the fact that the conjugate variables are the functions of the variables $\tilde{X}^a$. The metric is quite identical in form as the previous one (see \eqref{new2}). Here also, by the earlier argument, the nature of divergence of the Ricci scalar on $P_{VV}$ is $R_2^{(PV)}\sim\mathcal{O}(1/P_{VV}^2)$. 
When the van der Walls system reaches to the critical point, then the condition of the point of inflection gives $(\partial^2 P)/(\partial V^2)=P_{VV}=0$, implying the Ricci scalar for this metric diverges.

In the above, we have presented the explicit form of two Legendre invariant thermogeometrical metrics: One is in $F$-picture and the other one is in the $P$-picture. The first one describes the condition $\partial P/\partial V=0$ as the divergence of the corresponding Ricci-scalar while the other one gives the same result for $\partial^2P/\partial V^2=0$. Hence, we conclude that the van der Waals like critical point of a black hole can be equivalently analyzed by properly defined thermogeometry.

Let us now make a comment on the above analysis. When we are saying that the Ricci scalar diverges at the critical point, we are assuming that the numerator does not vanish at this point in all cases. If such situation arises, then one has to be careful. Since it is $0/0$ form, may be L'Hospital's rule can help to conclude. We shall take an explicit case where such situation would not arise. Let us discuss for the simplest example. Consider the Reissner--Nordstrom AdS (RNAdS) metric:
\begin{equation}
ds^2 =-\tilde{f}(r)dt^2+\frac{dr^2}{\tilde{f}(r)}+r^2d\Omega^2~, \label{RNADS}
\end{equation}
with $\tilde{f}(r)=1-(2M/r)+(Q^2/r^2)+(r^2/l^2)$ where, $M$, $Q$ and $l$ being the mass, the electric charge ($\equiv Y$) and the AdS curvature radius (related to the cosmological constant as $\Lambda=-3/l^2$) respectively. The conjugate variables for the invariant $\theta_F$ picture can be written in terms of the thermodynamic variables as
\begin{align}
P=\frac{T}{2CV^{\frac{1}{3}}}-\frac{1}{8\pi C^2V^{\frac{2}{3}}}+\frac{Q^2}{8\pi C^4V^{\frac{4}{3}}}~,
\nonumber
\\
S=\pi C^2 V^{\frac{2}{3}}~,  \ \ \ \ \ \ \phi=\frac{Q}{C V^{\frac{1}{3}}}~,
\nonumber
\\
F=\frac{CV^\frac{1}{3}}{2}-\pi TC^2V^\frac{2}{3}+\frac{Q^2V^{-\frac{1}{3}}}{2C}~, \label{VALUES}
\end{align}
with the relation of the event horizon radius $r_{+}=CV^{\frac{1}{3}}$, where $C=(3/4\pi)^{\frac{1}{3}}$. Since in this case $F_{TT}=0=F_{TQ}$, the thermogeometrical metric \eqref{new2} takes reduces to
\begin{align}
\mathcal{G}^{(PV)}_{1(RN)}=-f(V,Q)dV^2+A(V,Q)dQ^2~,
\end{align}
which is a two dimensional metric. Now, the calculation of the Ricci-scalar becomes very straight forward, which is given by the relation 
\begin{align}
R_{1(RN)}^{(PV)}=\frac{1}{2f^2A^2}[f(f_QA_Q-A_V^2)+A\{f_Q^2-f_VA_V
\nonumber
\\
-2f(f_{QQ}-A_{VV})\}]~. 
\label{RICCI1}
\end{align}
It is very obvious that the Ricci-scalar diverges as  $\mathcal{O}(1/f^2)$. But, one should verify whether the numerator vanishes at the critical point. The critical values of the thermodynamic quantities can be obtained satisfying the two criticality conditions (i.e., $F_{VV}=P_{VV}=0$). One can find the numerator (i.e., $A(f_Q^2-f_VA_V)$) of the Ricci-scalar gives small finite nonzero result for the critical values. But, not surprisingly, the total expression of the scalar curvature diverges at the critical point again.

For the geometrical description of the point of inflection of an RN-AdS black hole in the invariant $\theta_P$ picture the metric \eqref{G2} becomes
\begin{align}
\mathcal{G}^{(PV)}_{2(RN)}=-h(V,Q)dV^2+w(V,Q)dQ^2~,
\end{align}
with 
\begin{eqnarray}
&&h=P_{VV}(VP_V+TP_T+QP_Q)~;
\nonumber
\\
&&w=P_{QQ}(VP_V+TP_T+QP_Q)~. 
\end{eqnarray}
The above one is again a two-dimensional metric as the other metric coefficients containing $P_{TT}$ and $P_{TQ}$ vanish in this case. Calculation of the Ricci-scalar is again very straight forward yielding
\begin{align}
R_{2(RN)}^{(PV)}=\frac{1}{2h^2w^2}[h(h_Qw_Q-w_V^2)+w\{h_Q^2-h_Vw_V
\nonumber
\\
-2h(h_{QQ}-w_{VV})\}]~. 
\label{RICCI2}
\end{align}
Again the scalar curvature diverges as $\mathcal{O}(1/h^2)$. Here also, one can verify whether the numerator vanishes at the critical point. One can find that for the critical values, the numerator  of the Ricci-scalar (i.e., $w(h_Q^2-h_Vw_V)$) gives small finite nonzero result. But, the total expression of the scalar curvature diverges.

In this section, we have elaborately described the procedure to provide the geometric description of the black hole criticality at the extended phase space. Besides the $P$-$V$ criticality, as we have mentioned earlier, there are also the $T$-$S$ criticality and the $Y$-$X$ criticality in the non-extended phase space. In the following sections we shall find whether the previous arguments of the $P$-$V$ criticality in the extended phase space is also applicable for these two criticality of black holes in the non-extended phase space as well.

\section{The $T$-$S$ criticality}

Recently, there are many works that deals with the $T-S$ criticality \cite{Spallucci:2013osa, Mo:2016apo, Kuang:2016caz} of the black holes. The criticality, as said, comes due to the fact that both the $P-V$ and the $T-S$ space are dual. The expression of the pressure and the temperature comes from the same relation of black hole temperature \cite{Spallucci:2013osa} and the criticality conditions also looks alike both in $P-V$ and $T-S$ pictures.  In the $T-S$ phase space the conditions are $(\partial T/\partial S)_{Y_i}=0=(\partial^2 T/\partial S^2)_{Y_i}$.  In this case, one expresses $T$ as a function of $S$ and $Y_i$. We shall show here that our method, presented in the earlier section to give the geometrical description of the critical point, can also be used for this criticality as well. 

Let us first concentrate on the first condition. Here the relevant thermodynamic potential, as will be justified by the main analysis, is the internal energy $E(S, Y_i)$. The first law of the black hole ( in the non-extended phase space) is given as $dE=TdS+\sum_iX_idY_i$, where $E$ is identified as the black hole mass $M$. Hence, the thermodynamic variables are $\tilde{X}^a=\{S, Y_i\}$. Also, the conjugate variables are $\tilde{P}^a=\{T, X_i\}$ with $T=E_S$ and $X_i=E_{Y_i}$. The thermodynamic phase space $\mathcal{T}$ have the coordinates $Z^a=\{E, \tilde{X}^a, \tilde{P}^a\}$. The Legendre invariant metric, following the definition of \eqref{GENERAL}, can be written as 
\begin{align}
G^{(TS)}_1=\theta_E^2+(TS+\sum_iX_iY_i)(-dT dS+\sum_jdX_jdY_j),
\end{align}
 with the invariant $\theta_E=TdS+\sum_iX_idY_i$. Also, we assume a subspace $\mathcal{E}$ with the coordinates $\tilde{X}^a$. Let there be a smooth mapping $\varphi_E:\mathcal{E}\rightarrow\mathcal{T}$ and the subspace $\mathcal{E}$ is called the subspace of equilibrium states for the condition $\varphi^{*}_E(\theta_E)=0$. Now, the metric induced on $\mathcal{E}$ by means of $\mathcal{G}_1^{(TS)}=\varphi^{*}_E(G_1^{(TS)})$ can be written as
\begin{align} 
 \mathcal{G}^{(TS)}_1=-f'(S,Y_i)dS^2+\sum_{ij}g'_{ij}(S,Y_i)dY_idY_j,
\end{align} 
with
\begin{eqnarray}
&&f'(S,Y_i)=E_{SS}(SE_S+\sum_iY_iE_{Y_i})~;
\nonumber
\\
&&g'_{ij}(S,Y_i)=E_{Y_iY_j}(SE_S+\sum_kY_kE_{Y_k})~.
\end{eqnarray}
 For this multi-dimensional metric, if one calculates the scalar curvature it diverges at the critical point as $E_{SS}$ vanishes at that concerned point. Also, like the earlier case, the maximum divergence of the Ricci-scalar is proportional to the square inverse of $E_{SS}$ i.e., $R_1^{(TS)}|_{max. diver.} \sim \mathcal{O}(1/E_{SS}^2)~.$ For better understanding we assume the presence of the single charge rather then multiple ones. Taking this liberty, the metric gets the form: 
\begin{equation} 
\mathcal{G}^{(TS)}_1=-f'(S,Y)dS^2+g'(S,Y)dY^2~,
\end{equation}
 with $f'(S,Y)=E_{SS}(SE_S+YE_{Y})$ 
  and $g'(S,Y)=E_{YY}(SE_S+YE_{Y})$.
Again the scalar curvature can be calculated for this two dimensional thermogeometrical metric very straight-forwardly, which is given as-
\begin{align} 
  R^{(TS)}_1=\frac{1}{{2f'^2}g'^2}[f{'}(f{'}_Yg{'}_Y-g_S'^2)+g{'}(f{'}_Y^2-f{'}_Sg{'}_S
  \nonumber
  \\
  -2f{'}(f{'}_{YY}-g{'}_{SS}))].
\end{align} 
The above shows that $R^{(TS)}_1\sim\mathcal{O}(1/f'^2)$.
   Now $\partial T/\partial S=M_{SS}=0$ at the critical point. So, the Ricci-scalar diverges for this metric at the critical point. Thus, one condition gives the singular Ricci scalar at the critical point. Let us look at the other condition of the $T$-$S$ criticality.

As done earlier, to give the geometric interpretation of the second condition, one has to express the temperature as a function of the entropy $S$ and the charges $Y_i$, thereby obtaining 
$dT=(\partial T/\partial S)_{Y_i}dS+\sum_i(\partial T/\partial Y_i)_SdY_i=T_SdS+\sum_iT_{Y_i}dY_i$. 
 We define the thermodynamic phase space $\mathcal{T}$ with coordinates $Z^a=\{T, \tilde{X}^a, \tilde{P}^a\}$, with $\tilde{X}^a=\{S, Y_i\}$ being the thermodynamic variables and $\tilde{P}^a=\{T_S, T_{Y_i}\}$ being the corresponding conjugate variables. The thermogeometrical metric can be defined as 
\begin{align}
G^{(TS)}_2=\theta_T^2+(ST_S+\sum_iY_iT_{Y_i})(-dSdT_S+\sum_jdY_jdT_{Y_j}),
\end{align}
 which is Legendre invariant with $\theta_T=dT-T_SdS-\sum_iT_{Y_i}d{Y_i}$. Let there be the subspace $\mathcal{E}$ with a smooth mapping $\varphi_T:\mathcal{E}\rightarrow\mathcal{T}$ and the subspace $\mathcal{E}$ is called the space of equilibrium states for $\varphi^{*}_T(\theta_T)=0$. So, the metric induced on $\mathcal{E}$ by the relation $\mathcal{G}_2^{(TS)}=\varphi_T^{*}(G_2^{(TS)})$ can be written in the form
\begin{align} 
 \mathcal{G}^{(TS)}_2=-h{'}(S,Y_i)dS^2+\sum_{jk}k{'}_{jk}(S,Y_i)dY_jdY_k~, 
 \label{METTS}
\end{align} 
  with 
\begin{eqnarray}
&&h{'}(S, Y_i)=(ST_S+\sum_iY_iT_{Y_i})T_{SS}~;
\nonumber
\\
&&k{'}_{jk}(S,Y_i)=(ST_S+\sum_iY_iT_{Y_i})T_{Y_jY_k}~. 
\end{eqnarray}
The metric is identical to the previous one. Therefore, the earlier discussion is befitting in this case as well. So, the scalar curvature diverges at the critical point for this set up as well with the maximum divergence being proportional to the inverse square of $T_{SS}$, i.e., $R_2^{(TS)}|_{max. diver.} \sim \mathcal{O}(1/T_{SS}^2)~.$ 
  
   Thus we can see that the two criticality conditions in the $T$-$S$ picture of the non-extended phase space corresponds to the two singular scalar curvature. In other words, one can say that the $T$-$S$ criticality of the black holes can be obtained when the scalar curvature diverges simultaneously in the two pictures.
\section{$Y$-$X$ criticality}
Besides the $P-V$ and the $T-S$ criticality, for the black holes there are also the $Y-X$ criticality as shown in some recent works (for eg. \cite{Ma:2016aat}). This criticality is also described in the non-extended phase space and the conditions, like the earlier ones, are $(\partial Y_i/\partial X_i)=0=(\partial^2Y_i/\partial X^2_i)$. The appropriate quantity to start with is the Gibbs free energy to form the thermogeometrical metric for the first condition. The Gibbs free energy of the black holes is defined as $G=E-TS-\sum_iX_iY_i$ (see \cite{Cai:2004pz, Beauchesne:2012qk}), where $E=M$ in this case. Using the first law $dE=TdS+\sum_iX_idY_i$, one obtains $dG=-SdT-\sum_i Y_idX_i$. 
 The thermodynamic phase space $\mathcal{T}$ can be defined by the coordinates $Z^a=\{G, \tilde{X}^a, \tilde{P}^a\}$ with $\tilde{X}^a=\{T, X\}$ is the thermodynamic variables and $\tilde{P}^a=\{G_T=-S, G_{X_i}=-Y_i\}$ is the corresponding conjugate variables. Also, we assume the subspace $\mathcal{E}$ with the smooth mapping $\varphi_G:\mathcal{E}\rightarrow\mathcal{T}$ and $\mathcal{E}$ is called the space of equilibrium states for $\varphi_G^{*}(\theta_G)=0$. Like our previous cases, we want to define a thermogeometrical metric in a Legendre-invariant way (induced on $\mathcal{E}$ for $\varphi_G^{*}(\theta_G)=0$) which is given as 
 \begin{align}
 \mathcal{G}_1^{(Y X)}=-\sum_{ij}f{''}_{ij}(T, X_i)dX_idX_j+g{''}(T, X_i)dT^2~, 
 \label{MET1} 
 \end{align}
 with 
 \begin{eqnarray}
&&f{''}_{ij}(T, X_i)=G_{X_iX_j}(-TS-\sum_kY_kX_k)~;
\nonumber
\\
&&g{''}(T, X_i)=G_{TT}(-TS-\sum_kY_kX_k)~.
 \end{eqnarray}
 At the critical point, since $G_{X_iX_i} =-\partial Y_i/\partial X_i$, we have vanishing of $G_{X_iX_i}$. Therefore, like the previous cases, the scalar curvature corresponding to this metrics becomes singular and the maximum divergence of the scalar curvature is proportional to the inverse square of $G_{X_iX_i}$ i.e., $R_1^{(YX)}|_{max. diver.} \sim \mathcal{O}(1/G_{X_iX_i}^2)~.$ 
 
To give the geometric interpretation of the second condition we express the charge $Y_i$ as a function the existing potential $X_i$ and temperature T i.e., $dY_i=\sum_{j}Y_i{_{X_j}}dX_j+Y_i{_{T}}dT$ with $Y_i{_{X_j}}=(\partial Y_i/\partial X_j)_T$ and $Y_i{_{T}}=(\partial Y_i/\partial T)_{X_i}$. The thermodynamic phase space $\mathcal{T}$ is defined by the coordinates $Z^a=\{Y_i, \tilde{X}^a, \tilde{P}^a\}$ with $\tilde{X}^a=\{X_j, T\}$ and $\tilde{P}^a=\{Y_i{_{X_j}}, Y{_i}_T\}$. The Legendre-invariant metric is given as (which is induced on $\mathcal{E}$ and $\mathcal{E}$ is defined as the same way as done earlier)
\begin{align}
\mathcal{G}^{(YX)}_2=-\sum_{i,j}h{''}_{mij}(T, X_i)dX_idX_j+k{''}_m(T, X_i)dT^2, \label{MET22}
\end{align}
with 
\begin{eqnarray} 
&&h{''}_{mij}(T, X_i)=Y{_m}_{X_iX_j}(\sum_kY{_m}_{X_k}X_k+Y{_m}_TT)~;
\nonumber
\\
&&k{''}_m(T, X_i)=Y{_m}_{TT}(\sum_kY{_m}_{X_k}X_k+Y{_m}_TT)~.
\end{eqnarray}
Now, the metric elements $h{''}_{iii}(T,X_i)$ vanishes at the critical point as $Y{_i}_{X_iX_i}=0$ at that concerned point. Therefore, like the earlier cases, the scalar curvature diverges at that point with the maximum divergence being proportional to the inverse square of $Y{_i}_{X_iX_i}$ i.e., $R_2^{(YX)}|_{max. diver.} \sim \mathcal{O}(1/Y{_i}_{X_iX_i}^2)~.$ So, at the critical point, the condition $(\partial^2Y_i/\partial X_i^2)_T=0$ yields the divergence of the Ricci-scalar.

Thus we see, for each criticality we can obtain two sets of metrics and the corresponding Ricci-scalars. In every situation, the critical point is determined by the divergence of these scalars. For the $P-V$ criticality has been discussed in extended phase-space; whereas others are in usual phase-space. This is the consequence of the original description of criticalities in different situations.
\section{Conclusions}
The fact, that the black hole horizons have the thermodynamic structure, is now well established and is universally acknowledged. Even in our previous work \cite{Bhattacharya:2016kbm}, we have shown that the thermodynamic structure is still maintained for the time-dependent black holes as well, though lots of complexities arises into the analysis due to incorporating the time dependence into the theory. Apart from the idea that the governing equations of the black hole physics having a thermodynamic structure, physicists were keen to know since long whether all the facets of the conventional thermodynamics is followed by the black holes as well. One of the conspicuous outcome of the analysis resulted in the conclusion that the critical behavior of the usual thermodynamic system can also be found in the black hole mechanics as well. To fit the criticality properly in the black hole thermodynamics, it became important to describe the criticality in terms of the geometry. In this spirit, Weinhold and later Ruppeiner appeared with the idea of the formation of the thermogeometrical metric to describe the thermal interaction in terms of the curvature of the geometry. The two theory gives the inconsistent inference to each other which is believed to come due to the fact that those two proposed theory was not formed in a Legendre invariant way. Later, Quevedo et al. was able to formulate the thermogeometrical metric in a Legendre invariant way to describe the phase transition in a geometrical way. In the study of the black hole criticality, there are two distinct approaches: one approach is concerned about the behavior of the heat capacity and the inverse of the  isothermal compressibility, the two quantities which diverges at the critical point. The other one discusses the phase transition in the AdS background with the role of the cosmological constant as pressure in the extended phase space. Unlike the usual thermodynamic cases, for the black holes one can find P-V criticality, T-S criticality and also the Y-X criticality as mentioned earlier. There had been made multiple isolated attempts to explain those criticality in a geometrical way. Those attempts mainly highlights the criticality as the point of diverging heat capacity rather then one as a point of inflection and so far, as we know, no work could explain all these criticality in a unified way. 

In this work, we have given the geometrical description of the critical point as a point of inflection and the whole work is independent of any particular spacetime. Moreover, the analysis can explain all the criticalities in a unified way. Besides, we have followed the recent trend of defining the metric in a Legendre invariant way to avoid the inconsistency in the result. For the P-V criticality, we have taken the Helmholtz free energy (F) in the extended phase space and have defined our thermogeometrical metric in the invariant $\theta_F$ picture. It has been shown that the scalar curvature diverges at the critical point while satisfying one criticality condition i.e. $(\partial P/\partial V)_{T,Y}=0$. For the another condition, i.e., the condition of the point of inflection, the pressure is expressed as a function of volume, temperature and the charges and the metric has been defined in invariant $\theta_P$ picture. In this case as well one gets the singular behavior of the scalar curvature at the critical point. Later, it has been shown that the previous discussions can further be extended for the other criticalities (T-S and Y-X) as well. For the T-S criticality, the Helmholtz free energy is replaced by the internal energy (E) in the non-extended phase space and the metric is defined in invariant $\theta_E$ picture to get the singular behavior of the thermogeometrical metric at the critical point. The second metric for this criticality is made on the invariant $\theta_T$ picture. It has been shown that the scalar curvature in that picture also diverges at the critical point of the T-S criticality. The same procedure has been followed for the Y-X criticality as well where the two metrics were formulated in $\theta_G$ (G being the Gibbs free energy) and $\theta_Y$ picture and the Ricci-scalar corresponding to these metrics diverges at the critical point.

Here we want to mention the fact that in the most of the works people takes the thermodynamic potential as the function of the extensive variables to form the thermogeometrical metric. For the ordinary system in the thermodynamics, the fundamental equation comes from the internal energy which is the function of the extensive variables only. But, for the non-ordinary system such as black holes, the fundamental equation might come from the thermodynamic potential which is a function of both the extensive and the intensive variables as we have seen here. Although, \cite{Quevedo:2007mj} only discusses this issue in a brief, but it has also taken the thermodynamic potential as the function of only the extensive variables. Also, \cite{Quevedo:2008xn} takes the thermodynamic potential as the function of both the extensive and the intensive variables (see eq. 6.14) {\footnote{Also we find \cite{Cai:1998ep}, where the Ruppeiner metric is written in T,J coordinates (see eqn. (21))}}. We want to mention it clearly that taking the thermodynamic potential as the function of the extensive variables has nothing to do with the Legendre-invariance. Moreover, apart from defining the thermogeometrical metric in the thermodynamic potential picture (such as in $F$, $E$, $G$ etc.), one can also define it in thermodynamic variable picture (such as in S picture which is done in \cite{Alvarez:2008wa, Quevedo:2007mj}). Here, we have done the same to describe the point of inflection. But unlike in S-picture, the thermodynamic variables in our cases are both the intensive and the extensive variables. Again, it has no connection with the Legendre invariance, the only condition which we have given the utmost importance in our analysis.

Let us make some additional comments. In the $T-S$ and the $Y-X$ cases, the critical point is determined by the conditions in which $\Lambda=-3/l^2$ is kept constant (where $l$ is the AdS curvature radius). Therefore, one  is not needed to go to the extended phase space for these cases. The usual first law is well enough to describe the criticality.
 However, it should be remembered that this criticality can be described only for the AdS black holes and not for the asymptotically flat black holes as the critical temperature at the asymptotic flat limit ($l\rightarrow\infty$) vanishes and the charge and the entropy diverges (for an explicit example see Eq.(3.14) of \cite{Kuang:2016caz}). Therefore, the AdS nature of the black hole is required.
 On the other hand, the arguments made in the sections \ref{SEC1ST} and \ref{SEC2ND} are not applicable for the asymptotically flat black holes. The reason is in these cases cosmological constant $\Lambda=0$ and hence there is no pressure term in first law of thermodynamics (note pressure is defined by the relation $P=-\Lambda/8\pi$). Due to the absence of the pressure term, the question of presence of the $P-V$ criticality does not arise.

Another very interesting point may be worth mentioning. There are some works \cite{Monteiro:2009tc, Schiffrin:2013zta} which discuss the thermodynamic instability of the black holes and also argues how the thermodynamic instability is connected to the dynamical instability. For the normal (non-extended) phase space, there is a prescription which says that the thermodynamic and dynamical instability implies $\mathcal{B}<0$ under certain conditions (for details and the definition of $\mathcal{B}$ see \cite{Schiffrin:2013zta}). The quantity $\mathcal{B}$ is called ``canonical energy''. Now it has been proved that if there exists a turning point (see \cite{Schiffrin:2013zta} for the definition) in the system, there must be a thermodynamic instability on one side of that turning point. For the present case, note that the critical point is not an extrema rather a point of inflection and, hence, it is not a turning point. Therefore, we can not apply this theorem  to discuss whether there exists any thermodynamic instability in the current scenario. So one needs to go back to the basic condition (i.e. the condition on $\mathcal{B}$) to explore such possibilities. Now remember that the $P-V$ criticality appears in the extended phase space. Therefore in this case, one needs to construct the proper quantity like $\mathcal{B}$ from the first principle; i.e. one is needed to start from the scratch and all these prescriptions should be revamped. Of course, the cases like $T-S$ and $Y-X$ can be discussed in the already existing framework by explicit use of the particular black hole metric. Considering it to be a wonderful and sizable future work, we have kept aside the discussion of the dynamic and the thermodynamic stability for the future and have focused ourselves for the geometrical description of the critical point in a metric independent way.

To sum up, we estimate the critical point as a point where the Ricci-scalar of the two different pictures diverge simultaneously. The importance of our analysis lies on the novelty in the ideas which is discussed throughout the paper. Most importantly, we have geometrically described the critical point in terms of the point of inflection which is completely new. The analysis must give more inputs in both van der Waals like phase transition for AdS black holes as well as geometrothermodynamics. Hope we shall provide more insights later in this regards.

\vskip 4mm
{\section*{Acknowledgments}}
\noindent  
We want to give thanks to the anonymous referee for pointing out some important issues which helped to improve the earlier version.
The research of one of the authors (BRM) is supported by a START-UP RESEARCH GRANT (No. SG/PHY/P/BRM/01) from Indian Institute of Technology
Guwahati, India.

\vskip 4mm
\appendix
\section{Proof of the first law for AdS black holes with varying cosmological constant} \label{APPEN1}
The first law of black hole mechanics for a AdS black hole with varying cosmological constant has been demanded in the earlier works in the form mentioned in \eqref{1ST}.
Most of the earlier works does not provide a rigorous proof of this law.
In the above mentioned thermodynamic relation, $M$ is the black hole mass which is interpreted as the Enthalpy of the system. Also, $X=\{\Omega, \phi\}$ and $Y=\{J, Q\}$ with $\Omega$, $\Phi$, $J$ and $Q$ being the angular velocity, electric potential, angular momentum and electric charge (that includes all the charges due to the symmetry  as well as the hair in the system) respectively.  Following is the rigorous proof of the earlier demand. For the simplicity of the calculation we shall  not incorporate the charge in the proof of the first law.

The action of a AdS black hole with varying cosmological constant reads
\begin{equation}
A=\int d^4x \sqrt{-g}L=\frac{1}{16\pi}\int\sqrt{-g}(R-2\Lambda). \label{ACTION}
\end{equation}
 To obtain the equation of motion and the boundary term one has to take the variation of the total action. Here it leads to the result
 \begin{align}
 \delta(\sqrt{-g}L)=\frac{1}{16\pi}[\sqrt{-g}(G_{ab}+\Lambda g_{ab})\delta g^{ab}
\nonumber
\\
+\sqrt{-g}\nabla_a\delta v^a-2\sqrt{-g}\delta\Lambda], \label{VARIATION}
 \end{align}
 where $G_{ab}=R_{ab}-\frac{1}{2}Rg_{ab}$ is the well known Einstein tensor and $\delta v^a=2P^{ibad}\nabla_b(\delta g_{id})$ with $P^{abcd}=\partial R/\partial R_{abcd}=\frac{1}{2}(g^{ac}g^{bd}-g^{ad}g^{bc})$. For the diffeomorphism symmetry $x^a\rightarrow x^a+\xi^a$ the operation of $\delta$ is replaced by the Lie-derivative $\pounds_{\xi}$, with $\pounds_{\xi}g^{ab}=-(\nabla^a\xi^b+\nabla^b\xi^a)$. Using the Bianchi identity $\nabla_aG^a_b=0$, one can obtain
 \begin{align}
 \pounds_{\xi}(\sqrt{-g}L)=\frac{\sqrt{-g}}{16\pi}[-2\nabla_a(G^a_b\xi^b)+\nabla_a\pounds_{\xi} v^a-2\nabla_a(\xi^a\Lambda)]~, \label{VARDE}
 \end{align}
 where we have used $\Lambda g_{ab}\pounds_{\xi} g^{ab}-2\pounds_{\xi}\Lambda=-2\Lambda\nabla_a\xi^a-2\xi^a\nabla_a\Lambda=-2\nabla_a(\xi^a\Lambda)$ since $\Lambda$ is a scalar.
 Also, for the diffeomorphism symmetry, the left hand side is given by $\pounds_{\xi}(\sqrt{-g}L)=\sqrt{-g}\nabla_a(L\xi^a)$. Therefore, finally one obtains 
 \begin{align}
 \sqrt{-g}\nabla_a[L\xi^a+\frac{1}{16\pi}(2G^a_b\xi^b+2\Lambda\xi^a-\pounds_{\xi}v^a)]=0, \label{CON}
 \end{align}
 Where one can obtain the explicit expression $\pounds_{\xi}v^a=\nabla_b\nabla^a\xi^b+\square\xi^a-2\nabla_a\nabla_b\xi^b$.
 So, the conserved Noether current for the diffeomorphism symmetry can be identified as
 \begin{align}
 J^a=L\xi^a+\frac{1}{16\pi}(2G^a_b\xi^b+2\Lambda\xi^a-\pounds_{\xi}v^a). \label{JA}
\end{align}
Using the results $2G^a_b\xi^b+R\xi^a=2R^a_j\xi^j=2[\nabla_b\nabla^a\xi^b-\nabla^a\nabla_b\xi^b]$ and the earlier mentioned expression of $\pounds_{\xi}v^a$, the above equation (eq. \eqref{JA}) yields $J^a=\nabla_b J^{ab}=\frac{1}{16\pi}\nabla_b[\nabla^a\xi^b-\nabla^b\xi^a]$, thereby obtaining the anti-symmetric Noether potential as
\begin{align}
J^{ab}=\frac{1}{16\pi}[\nabla^a\xi^b-\nabla^b\xi^a]. \label{JAB}
\end{align}
 Note, although the Noether current $J^a$ depends on the cosmological constant $\Lambda$, the Noether potential is independent of it. Also notice that the expression of the Noether potential does not change when one takes $\Lambda$ as a pure constant or even when one does not take $\Lambda$ in the theory. Later we shall show that the entropy and the energy of the black hole system directly related to the Noether potential. Therefore, it can be concluded that those quantities are exempted from any change due to  $\Lambda$. Also notice that to obtain the above expression we never took the help of Einstein's equations of motion. So it is an {\it off-shell} result.

On-shell one finds $G^a_b\xi^b=-\Lambda\xi^a$ from the equation of motion $G_{ab}+\Lambda g_{ab}=0$, yielding the on-shell Noether current (from \eqref{JA}) as 
\begin{align}
J^a=L\xi^a-\frac{\pounds_{\xi}v^a}{16\pi}, \label{JAON}
\end{align}
which implies
\begin{align}
\delta(\sqrt{-g}J^a)=\delta(\sqrt{-g}L)\xi^a+\sqrt{-g}L\delta\xi^a-\frac{\delta(\sqrt{-g}\pounds_{\xi}v^a)}{16\pi} \label{VARJ}
\end{align}
Let $\delta$ represents arbitrary field variation (here the variation of the metric tensor $g^{ab}$ and the cosmological constant $\Lambda$) that does not affect the vector $\xi^a$. So, $\delta\xi^a=0$, but $\delta\xi_a\neq 0$, then using \eqref{VARIATION} we obtain
\begin{align}
\delta(\sqrt{-g}J^a)=\frac{\xi^a}{16\pi}[\sqrt{-g}(G_{ij}+\Lambda g_{ij})\delta g^{ij}
\nonumber
\\
+\sqrt{-g}\nabla_i\delta v^i-2\sqrt{-g}\delta\Lambda]-\frac{\delta(\sqrt{-g}\pounds_{\xi}v^a)}{16\pi} \label{VAR3}
\end{align}
Let us calculate the above variation on-shell. which gives
\begin{align}
\delta(\sqrt{-g}J^a)=\frac{\sqrt{-g}\xi^a}{16\pi}[\nabla_i\delta v^i-2\delta\Lambda]-\frac{\delta(\sqrt{-g}\pounds_{\xi}v^a)}{16\pi} \label{VAR4}
\end{align}
Using straight forward calculation one can obtain $\pounds_{\xi}(\sqrt{-g}\delta v^a)=-2\sqrt{-g}\nabla_b(\xi^{[a}\delta v^{b]})+\sqrt{-g}\xi^a\nabla_b\delta v^b$ with $A^{[a}B^{b]}=\frac{1}{2}(A^aB^b-A^bB^a)$. Using this relation in \eqref{VAR4} one obtains
\begin{align}
\delta(\sqrt{-g}J^a)=\frac{1}{16\pi}[\pounds_{\xi}(\sqrt{-g}\delta v^a)-\delta(\sqrt{-g}\pounds_{\xi}v^a)
\nonumber
\\
+2\sqrt{-g}\nabla_b(\xi^{[a}\delta v^{b]})-2\sqrt{-g}\xi^a\delta\Lambda]. \label{VAR5}
\end{align}
Let us now denote 
\begin{align}
\omega^a=-\frac{1}{16\pi}\pounds_{\xi}(\sqrt{-g}\delta v^a)+\frac{1}{16\pi}\delta(\sqrt{-g}\pounds_{\xi}v^a). \label{OMEGA1}
\end{align}
 The significance of $\omega^a$ will be explained later in our discussions. Therefore, using this convention one obtains 
\begin{align}
\delta(\sqrt{-g}J^a)=-\omega^a+\frac{2\sqrt{-g}}{16\pi}[\nabla_b(\xi^{[a}\delta v^{b]})-\xi^a\delta\Lambda]. \label{VAR6}
\end{align}
Which implies that
\begin{align}
\omega^a=-\delta(\sqrt{-g}J^a)+\frac{2\sqrt{-g}}{16\pi}[\nabla_b(\xi^{[a}\delta v^{b]})-\xi^a\delta\Lambda].  \label{OMEGA2}
\end{align}
To realize all the things properly let us take the refuge of the classical mechanics. From the classical calculations one can obtain $\delta L(q, \dot{q})=[(\frac{\partial L}{\partial q})-d_t(\frac{\partial L}{\partial \dot{q}})]\delta q +d_t[p\delta q]$. where the first term is the equation of motion, that vanishes on-shell and the last term is the temporal derivative of the boundary term, which let us denote as $v$. Let us now adopt the conventions 
\begin{align}
v(\delta q)=p\delta q,
\nonumber
\\
v(\dot{q})=p\dot{q}. \label{CONV}
\end{align}
Using this conventions the variation of the Hamiltonian can be written as: 
\begin{align}
\delta H(q,p)=\delta[p(d_tq)]-d_t[p(\delta q)]=\delta[v(\dot{q})]-d_t[v(\delta q)]. \label{VARH}
\end{align}
Let us now try to realize the physical significance of the term $\omega^a$. But before that, to compare $\omega^a$ with the classical term, let us take the one-to-one correspondence in the following way.

 Let the metric tensor $g^{ab}$ corresponds to $q$ in the classical mechanics. The arbitrary variation of the metric tensor $\delta g^{ab}$ corresponds to $\delta q$ and the Lie-derivative of the metric tensor $\pounds_{\xi}g^{ab}$ corresponds to $\dot{q}$. Following this convention one can write
 \begin{align}
 \sqrt{-g}\pounds_{\xi}v^a\equiv v(\dot{q}),
 \nonumber
 \\
 \sqrt{-g}\delta v^a\equiv v(\delta q). \label{CONV2}
 \end{align}
 So, using the above mentioned convention in \eqref{CONV2} if one compares \eqref{OMEGA1} and \eqref{VARH} one actually finds (apart from the normalization factor $\frac{1}{16\pi}$) $\omega^a$ corresponds to $\delta \mathcal{H}$, where $\mathcal{H}$ is the Hamiltonian density. So, variation of the total Hamiltonian is given as 
\begin{eqnarray}
&&\delta H[\xi]=\delta\int_c d\Sigma_a \frac{\omega^a}{\sqrt{-g}}
\nonumber
\\
&&=-\int_cd\Sigma_a\nabla_b(J^{ab})+\frac{2}{16\pi}\int_c d\Sigma_a[\nabla_b(\xi^{[a}\delta v^{b]})
\nonumber
\\
&&-\xi^a\delta\Lambda], \label{DELH}
\end{eqnarray} 
where $c$ is the Cauchy surface. To obtain the last step we use $J^a=\nabla_bJ^{ab}$ in \eqref{OMEGA2}, where $J^{ab}$ is the anti-symmetric Noether potential. Here $d\Sigma_a=n_a\sqrt{h}d^3y$ is the infinitesimal surface area of three dimensional hyper-surface with $h$ being the determinant of 3-metric and $n_a$ being the normal to the surface. Let us now replace the cosmological constant $\Lambda$ by the pressure. In case of AdS black hole, the pressure is identified as $P=-\Lambda/8\pi$. After converting the volume integral to the surface integral, one can write \eqref{DELH} as
\begin{eqnarray}
&&\delta H[\xi]=-\frac{1}{2}\delta\int_{\mathcal{H}} d\Sigma_{ab}J^{ab}+\frac{1}{2}\delta\int_{\partial c_{\infty}} d\Sigma_{ab}J^{ab}
\nonumber
\\
&&-\frac{1}{16\pi}\int_{\partial c_{\infty}} d\Sigma_{ab} \xi^{[a}\delta v^{b]}
+\delta P\int_c d\Sigma_a\xi^a~, 
\label{NFL}
\end{eqnarray}
 where the new surface integrations are to be done on the on a bifurcation surface $\mathcal{H}$ and at 2-dimensional boundary of $c$ at asymptotic infinity (i.e., $\partial c_{\infty}$). On the bifurcation surface $\mathcal{H}$, taking $\xi^a$ as a timelike Killing vector, one must has $\xi^a=0$. So no contribution comes from the term containing $\xi^{[a}\delta v^{b]}$.  In this situation $\delta H[\xi]=0$, although $H[\xi]$ might not be zero. Then the first term on the right hand side can be identified as $-\frac{\kappa}{2\pi}\delta S$ from the Wald's prescription with $\kappa$ as the surface gravity. The second and the third term as a whole contributes as $\delta M-\Omega_{H}\delta J$ (see \cite{Wald:1993nt} for rigorous discussion).
 
  Now, let us concentrate on the last term; i.e. $\int_c d\Sigma_a\xi^a$ of \eqref{NFL}. The mentioned integral can be written further as $\int_{\mathcal{H}} \sqrt{h}d^3y n_a\xi^a-\int_{\mathcal{\infty}} \sqrt{h}d^3y n_a\xi^a$ where the first term is calculated at the horizon and the second term is evaluated at the asymptotic boundary. One finds that the first integral gives a finite result, whereas, the second term appears as a diverging one. To remove this divergence one needs to adopt the regularization procedure. Usually in the literature there are two prescriptions. One is adding a counter term in the action such that its contribution removes the divergence. Another one is to use the background subtraction method. In this case the background contribution removes the divergence and we get a finite volume. The addition of the extra term can be justified due to the fact that one can always introduce a total derivative term along with the actual Lagrangian as the governing dynamics is unaltered or due to the fact that the Noether potential ($J^{ab}$) is not uniquely determined (one can include arbitrary anti-symmetric tensor with it, when the divergence of that arbitrary term vanishes). Therefore, one has the freedom to include a term in the Lagrangian or in the Noether potential such that the mentioned divergence at the infinity can be removed. Hence, considering the covariant definition of the volume $V$ as
\begin{align} 
  V=-\int_{\mathcal{H}} \sqrt{h}d^3y n_a\xi^a+\int_{\mathcal{\infty}} \sqrt{h}d^3y [n_a\xi^a-(n_a\xi^a)_{BG}]~, \label{VOL}
\end{align}  
   where, the term containing $(n_a\xi^a)_{BG}$ is considered as the ``background contribution'' to obliterate the divergence, we interpret the last term of (\ref{NFL}) as $-V\delta P$. This volume, in literature, is usually called the thermodynamic volume. If one calculates \eqref{VOL} for the example \eqref{RNADS} it reduces to $V=4\pi r_{+}^3/3$ which has been used in \eqref{VALUES}. The pressure is given purely by the consmological constant and, hence, here it is given by $P=-\Lambda/8\pi$. The similar prescriptions have been adopted in \cite{Kastor:2009wy, Couch:2016exn}.
  
   Therefore, from \eqref{NFL} one obtains the desired result 
\begin{equation}
\delta M=T\delta S+V\delta P +\Omega_{H}\delta J~, 
\label{1STLAWS}
\end{equation}
 as one can identify $T=\frac{\kappa}{2\pi}$ being the black hole temperature. Thus the relation obtained in this analysis is the first law for the AdS black hole with varying cosmological constant. Here we have not incorporated any hair in this theory for the simplicity of the calculations. As we mentioned earlier, inclusion of hair gives the well-known contribution and, ultimately, the final expression while considering all the hairs is given by \eqref{1ST} . Our analysis matches with our earlier demand in that equation. One important comment in this regard is that here  we shall interpret $M$ as the enthalpy rather then the internal energy of the system to compare the AdS black hole with the fluid system. Remember in the usual fluid system the enthalpy gives the measure of energy to create the system (i.e., the internal energy) added with the amount required to establish the pressure and the volume of that system. Thus the role of black hole mass is also changed in the description of the black hole thermodynamics in the extended phase space.
 
 The fact, that the Noether potential is independent of the cosmological constant, has earlier been studied in \cite{Hollands:2005wt}. But, in that case, the cosmological constant ($\Lambda$) has been taken as a proper constant and, consequently, it does not appear in the potential. But, in our case, we have taken the variation of the cosmological constant along with the metric tensor to derive the first-law of the AdS black holes in the extended phase space. We have shown although the Noether current depends on the cosmological constant, the potential does not depend on it even for the off-shell case. In this regard, the interesting point to be noted is that the expression of the entropy and the mass-energy in the extended phase space is identical to the non-extended-phase space as the expression of the Noether-potential is the same in both the cases. However, in our case, the mass is taken to be finite whereas in \cite{Hollands:2005wt} the conditions for its finiteness and conservation have been systematically discussed.
 
 {\underline{\it Note added}:} During the preparation of our final draft, a paper appeared in arXiv \cite{Hyun:2017nkb} which also discusses the derivation of the first law of thermodynamics for an AdS black hole with varying cosmological constant by the Lagrangian approach. The main difference of our analysis with that work is that we are following the Wald's formalism whereas it deals with the quasi-local Abbott-Deser-Tekin (ADT) formalism in the analysis. Moreover our current (\ref{JA}) as well as the potential (\ref{JAB}) are off-shell.

\begin{widetext}
\section{Expression for Ricci scalar} \label{APPEN2}
Consider the following metric:
\begin{equation}
ds^2 = -f(V,T,Y)dV^2+g(V,T,Y)dT^2+2h(V,T,Y)dTdY+A(V,T,Y)dY^2~.
\label{metric}
\end{equation}
The Ricci scalar is found out to be (calculated by Mathematica $11$)
\begin{eqnarray}
&&R=\frac{1}{2 f^2 \left(h^2-A g\right)^2}  
{\bf{\Big[}}\Big\{A \Big(g_Y^2+(A_T-2 h_Y) g_T\Big)
+h \Big(-g_Y (2 h_Y+A_T)
+A_Y g_T
+4 h_Y h_T-2 A_T h_T
\nonumber
\\
&&+2 h (g_{YY}-2 h_{TY}
+A_{TT})\Big)
+g\Big(A_T^2+A_Y (g_Y-2 h_T)
-2 A (g_{YY}-2 h_{TY}+A_{TT})\Big)\Big\} f^2
\nonumber
\\
&&+f[-4 (f_{TY}-h_{VV}) h^3+(-h_V^2+2 f_Y g_Y+2 A_T f_T+2 A f_{TT}
-3 A_V g_V-2 A g_{VV}) h^2
\nonumber
\\
&&-A (g_Y f_T
+2 h_T f_T
+f_Yg_T
-4 g_V h_V) h+A^2 (f_T g_T-g_V^2)
+g^2 \Big(-A_V^2+A_Y f_Y-2 A (f_{YY}-A_{VV})\Big)
\nonumber
\\
&&-g \Big\{2 (f_{TT}-g_{VV}) A^2+\Big(3 h_V^2-2 h_Y f_T
+A_T f_T+f_Y (g_Y-2 h_T)-4 h f_{TY}-A_V g_V
+4 h h_{VV}\Big) A
\nonumber
\\
&&+h \Big(f_Y (2 h_Y+A_T)+A_Y f_T
-4 A_V h_V-2 h (f_{YY}-A_{VV})\Big)\Big\}]
\nonumber
\\
&&+(A g-h^2) \Big(g (f_Y^2-A_V f_V)
+A (f_T^2-f_V g_V)+h(2 f_V h_V-2 f_Y f_T)\Big)\bf{\Big]}
\label{RM}
\end{eqnarray}
\end{widetext}



\begin{thebibliography}{99}

\bibitem{Bekenstein:1973ur} 
  J.~D.~Bekenstein,
  ``Black holes and entropy,''
  Phys.\ Rev.\ D {\bf 7}, 2333 (1973).
  
\bibitem{Hawking:1974sw} 
  S.~W.~Hawking,
  ``Particle Creation by Black Holes,''
  Commun.\ Math.\ Phys.\  {\bf 43}, 199 (1975)
  Erratum: [Commun.\ Math.\ Phys.\  {\bf 46}, 206 (1976)].
  
\bibitem{Bardeen:1973gs} 
  J.~M.~Bardeen, B.~Carter and S.~W.~Hawking,
  ``The Four laws of black hole mechanics,''
  Commun.\ Math.\ Phys.\  {\bf 31}, 161 (1973).
 
   
\bibitem{Davies:1989ey} 
  P.~C.~W.~Davies,
  ``Thermodynamic Phase Transitions of {Kerr-Newman} Black Holes in De Sitter Space,''
  Class.\ Quant.\ Grav.\  {\bf 6}, 1909 (1989).
  
\bibitem{Hawking:1982dh} 
  S.~W.~Hawking and D.~N.~Page,
  ``Thermodynamics of Black Holes in anti-De Sitter Space,''
  Commun.\ Math.\ Phys.\  {\bf 87}, 577 (1983).
  
\bibitem{Chamblin:1999tk} 
  A.~Chamblin, R.~Emparan, C.~V.~Johnson and R.~C.~Myers,
  ``Charged AdS black holes and catastrophic holography,''
  Phys.\ Rev.\ D {\bf 60}, 064018 (1999)
  [hep-th/9902170].
  
\bibitem{Chamblin:1999hg} 
  A.~Chamblin, R.~Emparan, C.~V.~Johnson and R.~C.~Myers,
  ``Holography, thermodynamics and fluctuations of charged AdS black holes,''
  Phys.\ Rev.\ D {\bf 60}, 104026 (1999)
  [hep-th/9904197].
  
\bibitem{Kastor:2009wy} 
  D.~Kastor, S.~Ray and J.~Traschen,
  ``Enthalpy and the Mechanics of AdS Black Holes,''
  Class.\ Quant.\ Grav.\  {\bf 26}, 195011 (2009)
  [arXiv:0904.2765 [hep-th]].
  
\bibitem{Dolan:2010ha} 
  B.~P.~Dolan,
  ``The cosmological constant and the black hole equation of state,''
  Class.\ Quant.\ Grav.\  {\bf 28}, 125020 (2011)
  [arXiv:1008.5023 [gr-qc]].
  
\bibitem{Dolan:2011xt} 
  B.~P.~Dolan,
  ``Pressure and volume in the first law of black hole thermodynamics,''
  Class.\ Quant.\ Grav.\  {\bf 28}, 235017 (2011)
  [arXiv:1106.6260 [gr-qc]].
  
\bibitem{Dolan:2011jm} 
  B.~P.~Dolan,
  ``Compressibility of rotating black holes,''
  Phys.\ Rev.\ D {\bf 84}, 127503 (2011)
  [arXiv:1109.0198 [gr-qc]].
  
\bibitem{Dolan:2012jh} 
  B.~P.~Dolan,
  ``Where is the PdV term in the fist law of black hole thermodynamics?,''
  arXiv:1209.1272 [gr-qc].
  
  
\bibitem{Azreg-Ainou:2014lua} 
  M.~Azreg-Aïnou,
  ``Charged de Sitter-like black holes: quintessence-dependent enthalpy and new extreme solutions,''
  Eur.\ Phys.\ J.\ C {\bf 75}, no. 1, 34 (2015)
  [arXiv:1410.1737 [gr-qc]].

  
\bibitem{Azreg-Ainou:2014twa} 
  M.~Azreg-Aïnou,
  ``Black hole thermodynamics: No inconsistency via the inclusion of the missing $P-V$ terms,''
  Phys.\ Rev.\ D {\bf 91}, 064049 (2015)
  [arXiv:1411.2386 [gr-qc]].
  
  
\bibitem{Banerjee:2011cz} 
  R.~Banerjee and D.~Roychowdhury,
  ``Critical phenomena in Born-Infeld AdS black holes,''
  Phys.\ Rev.\ D {\bf 85}, 044040 (2012)
  [arXiv:1111.0147 [gr-qc]].
  
\bibitem{Banerjee:2012zm} 
  R.~Banerjee and D.~Roychowdhury,
  ``Critical behavior of Born Infeld AdS black holes in higher dimensions,''
  Phys.\ Rev.\ D {\bf 85}, 104043 (2012)
  [arXiv:1203.0118 [gr-qc]].
  
\bibitem{Majhi:2012fz} 
  B.~R.~Majhi and D.~Roychowdhury,
  ``Phase transition and scaling behavior of topological charged black holes in Horava-Lifshitz gravity,''
  Class.\ Quant.\ Grav.\  {\bf 29}, 245012 (2012)
  [arXiv:1205.0146 [gr-qc]].
  
\bibitem{Lala:2012jp} 
  A.~Lala,
  ``Critical phenomena in higher curvature charged AdS black holes,''
  Adv.\ High Energy Phys.\  {\bf 2013}, 918490 (2013)
  [arXiv:1205.6121 [gr-qc]].
  
\bibitem{Ma:2014tka} 
  M.~S.~Ma, F.~Liu and R.~Zhao,
  ``Continuous phase transition and critical behaviors of 3D black hole with torsion,''
  Class.\ Quant.\ Grav.\  {\bf 31}, 095001 (2014)
  [arXiv:1403.0449 [gr-qc]].
  
\bibitem{Azreg-Ainou:2014gja} 
  M.~Azreg-Aïnou, G.~T.~Marques and M.~E.~Rodrigues,
  ``Phantom black holes and critical phenomena,''
  JCAP {\bf 1407}, 036 (2014)
  [arXiv:1405.5745 [gr-qc]].
  
\bibitem{Liu:2013koa} 
  J.~X.~Mo and W.~B.~Liu,
  ``Phase transitions, geometrothermodynamics and critical exponents of black holes with conformal anomaly,''
  Adv.\ High Energy Phys.\  {\bf 2014}, 739454 (2014)
  [arXiv:1312.0679 [hep-th]].
  
\bibitem{Kubiznak:2012wp} 
  D.~Kubiznak and R.~B.~Mann,
  ``P-V criticality of charged AdS black holes,''
  JHEP {\bf 1207}, 033 (2012)
  [arXiv:1205.0559 [hep-th]].
  
\bibitem{Kubiznak:2016qmn} 
  D.~Kubiznak, R.~B.~Mann and M.~Teo,
  ``Black hole chemistry: thermodynamics with Lambda,''
  arXiv:1608.06147 [hep-th].
  
\bibitem{Mandal:2016anc} 
  A.~Mandal, S.~Samanta and B.~R.~Majhi,
  ``Phase transition and critical phenomena of black holes: A general approach,''
  Phys.\ Rev.\ D {\bf 94}, no. 6, 064069 (2016)
  [arXiv:1608.04176 [gr-qc]].
  
\bibitem{Majhi:2016txt} 
  B.~R.~Majhi and S.~Samanta,
  ``P-V criticality of AdS black holes in a general framework,''
  arXiv:1609.06224 [gr-qc].
  
\bibitem{Momeni:2016qfv} 
  D.~Momeni, M.~Faizal, K.~Myrzakulov and R.~Myrzakulov,
  ``Fidelity Susceptibility as Holographic PV-Criticality,''
  Phys.\ Lett.\ B {\bf 765}, 154 (2017)
  [arXiv:1604.06909 [hep-th]].
  
  \bibitem{WEINHOLD}
  F.~Weinhold,
  J.~Chem.~Phys.\  {\bf 63}, 2479 (1975);\  {\bf 63}, 2484 (1975);\  {\bf 63}, 2488 (1975);\  {\bf 63}, 2496 (1975);\  {\bf 65}, 559 (1976).
  
  \bibitem{RUPP}
  G.~Ruppeiner,
  Phys.\ Rev.\ A\ {\bf 20}, 1608 (1979).
  
\bibitem{Ruppeiner:1995zz} 
  G.~Ruppeiner,
  ``Riemannian geometry in thermodynamic fluctuation theory,''
  Rev.\ Mod.\ Phys.\  {\bf 67}, 605 (1995)
  Erratum: [Rev.\ Mod.\ Phys.\  {\bf 68}, 313 (1996)].

\bibitem{Quevedo:2006xk} 
  H.~Quevedo,
  ``Geometrothermodynamics,''
  J.\ Math.\ Phys.\  {\bf 48}, 013506 (2007)
  [physics/0604164].
  
\bibitem{Quevedo:2007mj} 
  H.~Quevedo,
  ``Geometrothermodynamics of black holes,''
  Gen.\ Rel.\ Grav.\  {\bf 40}, 971 (2008)
  [arXiv:0704.3102 [gr-qc]].
  
\bibitem{Quevedo:2008xn} 
  H.~Quevedo and A.~Sanchez,
  ``Geometrothermodynamics of asymptotically de Sitter black holes,''
  JHEP {\bf 0809}, 034 (2008)
  [arXiv:0805.3003 [hep-th]].
  
\bibitem{Quevedo:2008ry} 
  H.~Quevedo and A.~Sanchez,
  ``Geometric description of BTZ black holes thermodynamics,''
  Phys.\ Rev.\ D {\bf 79}, 024012 (2009)
  [arXiv:0811.2524 [gr-qc]].
  
\bibitem{Alvarez:2008wa} 
  J.~L.~Alvarez, H.~Quevedo and A.~Sanchez,
  ``Unified geometric description of black hole thermodynamics,''
  Phys.\ Rev.\ D {\bf 77}, 084004 (2008)
  [arXiv:0801.2279 [gr-qc]].
  
\bibitem{Quevedo:2011np} 
  H.~Quevedo and M.~N.~Quevedo,
  ``Fundamentals of Geometrothermodynamics,''
  arXiv:1111.5056 [math-ph].
  
\bibitem{Quevedo:2017tgz} 
  H.~Quevedo, M.~N.~Quevedo and A.~Sanchez,
  ``Homogeneity and thermodynamic identities in geometrothermodynamics,''
  arXiv:1701.06702 [gr-qc]. 
  
\bibitem{Quevedo:2016cge} 
  H.~Quevedo, M.~N.~Quevedo and A.~Sanchez,
  ``Geometrothermodynamics of phantom AdS black holes,''
  Eur.\ Phys.\ J.\ C {\bf 76}, no. 3, 110 (2016)
  [arXiv:1601.07120 [gr-qc]].
  
\bibitem{Quevedo:2016swn} 
  H.~Quevedo, M.~N.~Quevedo and A.~Sánchez,
  ``Einstein-Maxwell-dilaton phantom black holes: Thermodynamics and geometrothermodynamics,''
  Phys.\ Rev.\ D {\bf 94}, no. 2, 024057 (2016)
  [arXiv:1606.02048 [gr-qc]].
  
  
\bibitem{Banerjee:2016nse} 
  R.~Banerjee, B.~R.~Majhi and S.~Samanta,
  ``Thermogeometric phase transition in a unified framework,''
  Phys.\ Lett.\ B {\bf 767}, 25 (2017)
  [arXiv:1611.06701 [gr-qc]].
  
\bibitem{Spallucci:2013osa} 
  E.~Spallucci and A.~Smailagic,
  ``Maxwell's equal area law for charged Anti-deSitter black holes,''
  Phys.\ Lett.\ B {\bf 723}, 436 (2013)
  [arXiv:1305.3379 [hep-th]].
  
\bibitem{Kuang:2016caz} 
  X.~M.~Kuang and O.~Miskovic,
  ``Thermal phase transitions of dimensionally continued AdS black holes,''
  arXiv:1611.10194 [hep-th].
  
\bibitem{Shen:2005nu} 
  J.~y.~Shen, R.~G.~Cai, B.~Wang and R.~K.~Su,
  ``Thermodynamic geometry and critical behavior of black holes,''
  Int.\ J.\ Mod.\ Phys.\ A {\bf 22}, 11 (2007)
  [gr-qc/0512035].
  
\bibitem{Zhang:2015ova} 
  J.~L.~Zhang, R.~G.~Cai and H.~Yu,
  ``Phase transition and thermodynamical geometry of Reissner-Nordström-AdS black holes in extended phase space,''
  Phys.\ Rev.\ D {\bf 91}, no. 4, 044028 (2015)
  [arXiv:1502.01428 [hep-th]].
  
  
\bibitem{Mo:2016apo} 
  J.~X.~Mo, G.~Q.~Li and Y.~C.~Wu,
  ``A consistent and unified picture for critical phenomena of $f(R)$ AdS black holes,''
  JCAP {\bf 1604}, no. 04, 045 (2016)
  [arXiv:1602.01251 [gr-qc]].
  
\bibitem{Zeng:2015tfj} 
  X.~X.~Zeng, H.~Zhang and L.~F.~Li,
  ``Phase transition of holographic entanglement entropy in massive gravity,''
  Phys.\ Lett.\ B {\bf 756}, 170 (2016)
  [arXiv:1511.00383 [gr-qc]].
  
\bibitem{Zeng:2015wtt} 
  X.~X.~Zeng and L.~F.~Li,
  ``Van der Waals phase transition in the framework of holography,''
  Phys.\ Lett.\ B {\bf 764}, 100 (2017)
  [arXiv:1512.08855 [hep-th]].
  
\bibitem{Ma:2016aat} 
  Y.~B.~Ma, R.~Zhao and S.~Cao,
  ``Q- $\Phi $ criticality in the extended phase space of $(n+1)$ -dimensional RN-AdS black holes,''
  Eur.\ Phys.\ J.\ C {\bf 76}, no. 12, 669 (2016)
  [arXiv:1607.00793 [hep-th]].
  
\bibitem{Hendi:2015eca}
  S.~H.~Hendi, S.~Panahiyan, B.~Eslam Panah and M.~Momennia,
 ``Phase transition of charged black holes in massive gravity through new methods,''
  Annalen Phys.\  {\bf 528}, no. 11-12, 819 (2016)
  [arXiv:1506.07262 [hep-th]].
  
\bibitem{Hendi:2015rja} 
 S.~H.~Hendi, S.~Panahiyan, B.~Eslam Panah and M.~Momennia,
  ``A new approach toward geometrical concept of black hole thermodynamics,''
  Eur.\ Phys.\ J.\ C {\bf 75}, no. 10, 507 (2015)
  [arXiv:1506.08092 [gr-qc]].
  
  
  
\bibitem{Baldiotti:2017ywq} 
  M.~C.~Baldiotti, R.~Fresneda and C.~Molina,
  ``A Hamiltonian approach for the Thermodynamics of AdS black holes,''
  arXiv:1701.01119 [hep-th].
  
\bibitem{Couch:2016exn} 
  J.~Couch, W.~Fischler and P.~H.~Nguyen,
  ``Noether charge, black hole volume and complexity,''
  arXiv:1610.02038 [hep-th].
  
\bibitem{Wald:1993nt} 
  R.~M.~Wald,
  ``Black hole entropy is the Noether charge,''
  Phys.\ Rev.\ D {\bf 48}, no. 8, R3427 (1993)
  [gr-qc/9307038].
  
  \bibitem{Hermann}
  R.~Hermann,
   ``Geometry, physics and systems,'' 
   Marcel Dekker, New York, 1973.
   
   \bibitem{Mrugala1}
   R.~Mrugala,
    ``Geometrical formulation of equilibrium phenomenological thermodynamics,''
     Rep.~Math.~Phys. {\bf 14}, 419 (1978).
     
   \bibitem{Mrugala2}
   R.~Mrugala, 
   ``Submanifolds in the thermodynamic phase space,''
    Rep.~Math.~Phys. {\bf 21}, 197 (1985).
  
\bibitem{Cai:2004pz} 
  R.~G.~Cai and A.~Wang,
  ``Thermodynamics and stability of hyperbolic charged black holes,''
  Phys.\ Rev.\ D {\bf 70}, 064013 (2004)
  [hep-th/0406057].
  
\bibitem{Beauchesne:2012qk} 
  H.~Beauchesne and A.~Edery,
  ``Black hole free energy during charged collapse: a numerical study,''
  JHEP {\bf 1205}, 146 (2012)
  [arXiv:1203.2279 [gr-qc]].
  
\bibitem{Bhattacharya:2016kbm} 
  K.~Bhattacharya and B.~R.~Majhi,
  ``Temperature and thermodynamic structure of Einstein’s equations for a cosmological black hole,''
  Phys.\ Rev.\ D {\bf 94}, no. 2, 024033 (2016)
  [arXiv:1602.07879 [gr-qc]].
  
\bibitem{Cai:1998ep} 
  R.~G.~Cai and J.~H.~Cho,
  ``Thermodynamic curvature of the BTZ black hole,''
  Phys.\ Rev.\ D {\bf 60}, 067502 (1999)
  [hep-th/9803261].
  
\bibitem{Monteiro:2009tc} 
  R.~Monteiro, M.~J.~Perry and J.~E.~Santos,
  ``Thermodynamic instability of rotating black holes,''
  Phys.\ Rev.\ D {\bf 80}, 024041 (2009)
  [arXiv:0903.3256 [gr-qc]].
  
\bibitem{Schiffrin:2013zta} 
  J.~S.~Schiffrin and R.~M.~Wald,
  ``Turning Point Instabilities for Relativistic Stars and Black Holes,''
  Class.\ Quant.\ Grav.\  {\bf 31}, 035024 (2014)
  [arXiv:1310.5117 [gr-qc]].
  
\bibitem{Hollands:2005wt} 
  S.~Hollands, A.~Ishibashi and D.~Marolf,
  ``Comparison between various notions of conserved charges in asymptotically AdS-spacetimes,''
  Class.\ Quant.\ Grav.\  {\bf 22}, 2881 (2005)
  [hep-th/0503045].
  
\bibitem{Hyun:2017nkb} 
  S.~Hyun, J.~Jeong, S.~A.~Park and S.~H.~Yi,
  ``Thermodynamic Volume and the Extended Smarr Relation,''
  arXiv:1702.06629 [hep-th].
  

  

  

  

  
 




  
  
  

  

  
  
  
  



\end{thebibliography}
\end{document}